\documentclass[aps,prl,twocolumn,showpacs]{revtex4}
\usepackage{graphicx}
\newcommand{\bm}[1]{\mbox{\boldmath{$#1$}}}

\begin{document}

\title{Fully Band Resolved Scattering Rate in MgB$_2$ Revealed by
Nonlinear Hall Effect and Magnetoresistance Measurements}

\author{Huan Yang$^1$, Yi Liu$^1$, Chenggang Zhuang$^{2,3,4}$, Junren Shi$^{1,*}$, Yugui Yao$^1$, Sandro Massidda$^5$, Marco Monni$^5$, Ying
Jia$^1$, Xiaoxing Xi$^{3,4}$, Qi Li$^3$, Zi-Kui Liu$^4$, Qingrong
Feng$^{2}$, and Hai-Hu Wen$^{1,*}$}

\affiliation{$^1$Institute of Physics and Beijing National
Laboratory for Condensed Matter Physics, Chinese Academy of
Sciences, P.~O.~Box 603, Beijing 100080, P.~R.~China}

\affiliation{$^2$Department of Physics, Peking University, Beijing
100871, P.~R.~China}

\affiliation{$^3$Department of Physics, The Pennsylvania State
University, University Park, Pennsylvania 16802, USA}

\affiliation{$^4$Department of Materials Science and Engineering,
The Pennsylvania State University, University Park, Pennsylvania
16802, USA}

\affiliation{$^5$SLACS-INFM and Dipartimento di Fisica,
Universit\`{a} di Cagliari, I-09042 Monserrato (Ca), Italy}

\date{\today}

\begin{abstract}

We have measured the normal state temperature dependence of the Hall
effect and magnetoresistance in epitaxial MgB$_2$ thin films with
variable disorders characterized by the residual resistance ratio
$RRR$ ranging from 4.0 to 33.3. A strong nonlinearity of the Hall
effect and magnetoresistance have been found in clean samples, and
they decrease gradually with the increase of disorders or
temperature. By fitting the data to the theoretical model based on
the Boltzmann equation and \emph{ab initio} calculations for a
four-band system, for the first time, we derived the scattering
rates of these four bands at different temperatures and magnitude of
disorders. Our method provides a unique way to derive these
important parameters in multiband systems.

\end{abstract}

\pacs{74.70.Ad, 71.15.Dx, 72.10.Di, 74.25.Fy}

\maketitle

The multiband character of MgB$_2$ \cite{review} dominates its
properties in both superconducting
\cite{coherence_length,twogap,MgB2IV,tunnelling,STMvortex} and
normal state \cite{MR_Mora,MR_Pallecchi,LiQ,Mazin}. In MgB$_2$,
there are two holelike quasi-two-dimensional $\sigma$ bands (bonding
$\sigma_1$ and antibonding $\sigma_2$), an ``electronlike''
antibonding ($\pi_1$) and a ``holelike'' bonding ($\pi_2$)
three-dimensional $\pi$ band \cite{Kortus,deHaas}. The electron
scattering rates in each band and between different bands are the
most critical parameters dictating all aspects of the properties of
MgB$_2$ \cite{MR_Pallecchi,LiQ,Mazin,Gurevich}. The temperature
dependence of the electron scattering rates arises from
electron-phonon (\textit{e}-ph) coupling, and the strong
\textit{e}-ph coupling between the $E_\mathrm{2g}$ phonon mode and
the $\sigma$ bands are responsible for the high $T_\mathrm{c}$ in
MgB$_2$ \cite{isotope}. Consequently, measuring the intraband and
interband scattering rates in MgB$_2$, with a goal to further
manipulate them in order to reveal new physics and achieve desirable
properties, has been central to many research studies. The
properties used to extract the band-resolved scattering rates
include electrical resistivity \cite{Mazin}, magnetoresistance (MR)
\cite{MR_Pallecchi,LiQ,Monni,JAP}, far-infrared spectroscopy
\cite{Ortolani}, and upper critical field $H_{c2}$\cite{Gurevich}.
For example, Monni {\it et al.} was able to derive the
temperature-dependent relaxation times, $\tau_\sigma$ and $\tau_\pi$
for the generalized $\sigma$ and $\pi$ bands, respectively, from the
MR measurement and {\it ab initio} calculations \cite{Monni}.
However, although the two $\sigma$ bands (or the two $\pi$ bands)
have similar properties, they are distinct from each other. For
example, the two $\pi$ bands have different types of carriers.
Without the knowledge of scattering rates in all the \emph{four}
different bands, the understanding of the multiband nature of
MgB$_2$ is incomplete. To our best knowledge, information from MR
alone is not sufficient to derive the scattering rates in four
different bands. In this Letter, we report results of strong
nonlinear Hall effect (NLHE) and large MR in pure epitaxial MgB$_2$
films. This made it possible to derive the scattering rates and
their dependencies on temperature and disorder in each of the four
bands, a significant advancement of the knowledge concerning the
multiband character of MgB$_2$.

The MgB$_2$ films used in this work were grown by the hybrid
physical-chemical vapor deposition (HPCVD) technique \cite{film} on
(0001) 6$H$-SiC substrates. They were epitaxial with a $c$-axis
orientation and the thickness was between 100 and 200$\;$nm. Films
with different magnitudes of disorder, thus different residual
resistivity, were grown at slightly different temperatures,
pressures, and growth rates. Nevertheless, all the films had similar
$T_\mathrm{c}$ around 40$\;$K and narrow x-ray diffraction rocking
curves. The longitudinal and the transverse resistivity were
measured with sweeping magnetic field at a fixed temperature or
sweeping temperature at a fixed field. All the measurements were
performed with magnetic field applied perpendicular to the $ab$
plane of the film. For the seven films investigated here, the
residual resistance ratios
$RRR\equiv\rho(300\;\mathrm{K})/\rho(41\;\mathrm{K})$ are 33.3,
24.5, 20.9, 14.4, 6.88, 6.4 and 4.0, and we mark them as RRR33.3,
RRR24.5, RRR20.9, RRR14.4, RRR9.9, RRR6.4, and RRR4.0, respectively.
Their corresponding residual resistivities $\rho_\mathrm{n}$ are
0.293, 0.347, 0.411, 0.740, 1.32, 2.55, and 4.66$\;\mu\Omega\,$cm,
respectively.

\begin{figure}
\includegraphics[width=6cm]{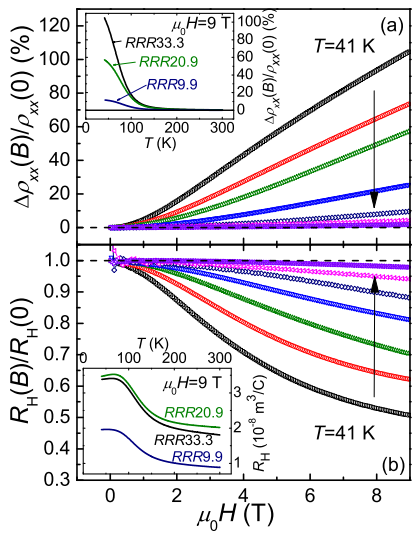}
\caption{(Color online) Field dependence of
$\Delta\rho_{xx}(B)/\rho_{xx}(0)$ (a), and $R_H(B)/R_H(0)$ (b) at
$T=41\;$K for seven samples with $RRR$ values from 33.3 to 4.0
(direction of the arrows). Insets in (a) and (b) shows the
temperature dependence of the normalized MR and $R_\mathrm{H}$ at
9$\;$T for sample RRR33.3, RRR20.9, and RRR9.9. One can see that the
MR and $R_\mathrm{H}$ decrease rapidly when the temperature is
increased. } \label{fig1}
\end{figure}

In Fig.~\ref{fig1}, we show the field dependence of
 $\Delta\rho_{xx}(B)/\rho_{xx}(0)$ and $R_\mathrm{H}(B)/R_\mathrm{H}(0)$ for
all the seven samples at $T=41\;$K. Here
$R_\mathrm{H}(B)=\rho_{xy}/B$ is the Hall coefficient at the
magnetic field $B$; $\rho_{xx}$ and $\rho_{xy}$ are the longitudinal
and transverse resistivity, and $\Delta \rho_{xx}(B)=
\rho_{xx}(B)-\rho_{xx}(0)$. The dashed lines show the cases for zero
MR and linear Hall effect, as one can expect in single band metals
with the spherical or the columned Fermi surface. MR and NLHE are
observed in all the samples, but the MR is larger and the NLHE is
stronger in cleaner samples. In sample RRR33.3 at 9$\;$T and
41$\;$K, a large MR of more than 100\% is observed, and the Hall
coefficient decreases to about half of the zero-field value. The
magnitude of MR is similar to an earlier report in a clean MgB$_2$
film due to the multiband effect.\cite{LiQ} For the dirty samples
with low $RRR$, because the interband scattering is very strong, the
multiband natures, such as the MR and the NLHE are weakened, and the
Kohler's rule is satisfied \cite{Budko}, which is not the case in
clean MgB$_2$ films.\cite{LiQ} The insets in Fig.~\ref{fig1} (a) and
Fig.~\ref{fig1} (b) present the temperature dependence of the
normalized MR and NLHE at 9$\;$T of the films RRR33.3, RRR20.9, and
RRR9.9. Both MR and the NLHE decrease rapidly with increasing $T$.
In a single band metal with anisotropic Fermi surface,
$R_\mathrm{H}$ should be very weakly $T$ dependent. However, as
shown in the inset of Fig.~\ref{fig1} (b), the Hall coefficient has
a strong $T$ dependence, especially for the clean samples. It should
be noted that the value of $R_\mathrm{H}$ is about an order of
magnitude larger than those reported by Eltsev {\it et al.} in
single crystals \cite{PRBEltsev}, which is not necessarily an
indication of a smaller density of charge carriers in this clean
film. As discussed below, due to the existence of multiple bands,
the Hall coefficient can no longer be written simply as $1/ne$ as in
single band materials.

\begin{figure}
\includegraphics[width=8cm]{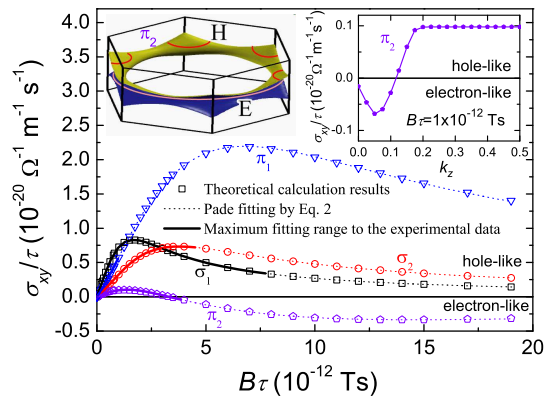}
\caption{(Color online) $B\tau$ dependence of $\sigma_{xy}/\tau$ for
the four bands. The left inset shows the Fermi surface of $\pi_2$
band, the electron-like orbit (`E') around the belly and the
hole-like orbit (`H') around the hills are indicated as the arc
lines. The right inset shows the partial Hall conductivity
$\sigma_{xy}^{(n)}(k_z)/\tau=\frac{e^2}{4\pi^3}\int
\mathrm{d}k_x\mathrm{d}k_y\int_{-k_z}^{k_z}\mathrm{d}k_z'v^{(n)}_{x}({\bf
k})\bar{v}_y^{(n)}(\mathbf{k})\left(-\frac{\partial
f}{\partial\varepsilon}\right)_{\varepsilon=\varepsilon_n(\mathbf{k})}$
contributed by orbits between $-k_z$ to $k_z$ of $\pi_2$ band as a
function of $k_z$. It demonstrates that the contribution from the
belly region is negative while the hill region is positive.}
\label{fig2}
\end{figure}

The large MR and NLHE originate directly from the multiband
character of MgB$_2$, from which we can extract information on the
electron scattering rates in the different bands. To do this, we
first need to determine the contribution of each band to the (Hall)
conductivity at the given electron scattering rate.  The complex
band structure of MgB$_2$ renders the simple formulae presented in
textbooks invalid.  We thus employ the semiclassical Boltzmann
theory with the relaxation time approximation~\cite{SolidState}. In this approximation,
the electron state is assumed to have a finite lifetime,
which is induced by all possible electron scattering processes,
including the intraband and inter-band scatterings. The conductivity tensor $\bm{\sigma}$ for the
$n^\mathrm{th}$ band reads:
\begin{equation} {\bm
\sigma}^{(n)}=e^2\int\frac{\mathrm{d}
\mathbf{k}}{4\pi^3}\tau_n {\bf v}_n({\bf
  k})\bar{{\bf v}}_n(\mathbf{k})\left(-\frac{\partial
  f}{\partial\varepsilon}\right)_{\varepsilon=\varepsilon_n(\mathbf{k})},
\end{equation}
where ${\bf
  v}_n(\mathbf{k})={\partial\varepsilon_n}/{\partial\hbar{\bf k}}$ is
the group velocity at the wave vector $\bf k$.  Under a magnetic
field, the wave vector of electron evolves by
$\hbar\dot{\mathbf{k}}_n=-(e/c){\bf v}_n({\bf k})\times{\bf B}$; and
$\bar{\mathbf{v}}_n(\mathbf{k})=\int_{-\infty}^{0}\mathrm{d}t\mathrm{e}^{t/\tau_n}{\bf
  v}_n({\bf k})/\tau_n$, is a weighted average of the velocity over
the past history of the electron orbit passing through $\bf k$; $f$
is the Fermi distribution function. Here, we have assumed that
$\tau_n$ is independent on the wave vector $\bf k$.  The band
structure necessary for the evaluation is determined from the {\it
ab initio} electronic structure calculations. Here we employed the
ultrasoft pseudopotential plane-wave method~\cite{calUPPW} with
generalized-gradient approximation~\cite{calGGA} for the exchange
and correlation potential. For the case of $B=0$, we obtain the
electronic density of state at the Fermi level and plasma frequency
$\omega_{xx}$ and $\omega_{zz}$ for each band in very good agreement
with those given by Liu {\it et al.}~\cite{Mazin}. For $B\neq0$,
$\bar{\bf v}_n$ is obtained by self-adaptive Runge-Kutta
integration. The calculated Hall conductivity
$\sigma_{xy}^{(n)}/\tau_n$, as a function of $B\tau_n$, is shown in
Fig.~\ref{fig2}.  The calculation results for both $\sigma_{xx}$ and
$\sigma_{xy}$ can be well interpolated by the Pad\'{e} series:
\begin{equation}
\begin{array}{l}\displaystyle{ \sigma_{xx}^{(n)}/\tau_n=\frac{a_1^{(n)}+a_2^{(n)}\cdot
(B\tau_n)^2}{1+b_{1}^{(n)}\cdot(B\tau_n)^2+b_{2}^{(n)}\cdot
(B\tau_n)^4}},\\\displaystyle{\sigma_{xy}^{(n)}/\tau_n=\frac{c_1^{(n)}\cdot
(B\tau_n) +c_2^{(n)}\cdot (B\tau_n)^3}{1+d_{1}^{(n)}\cdot(B\tau_n)^2+d_{2}^{(n)}\cdot
(B\tau_n)^4}},\end{array}\label{eq:fit}
\end{equation}
with the coefficients shown in Table~\ref{table1}.  The total
conductivity is the summation of the four bands. The \textit{ab
initio} calculation reveals some unexpected behaviors: (i) The
``electronlike'' $\pi_1$ band behaves as a hole band when the
magnetic field is normal to the $ab$ plane; (ii) The ``holelike''
band $\pi_2$ has a sign change from positive to negative with
increasing $B\tau$. The latter can be understood as a result of the
complex structure of the $\pi_2$ Fermi surface, which is composed of
electron-like belly and hole-like hills, as shown in the left inset
of Fig.~\ref{fig2}. At small $B\tau$, the conduction is dominated by
the small orbits around the hills, showing the hole-like behavior.
The opposite happens when $B\tau$ is large. These unusual behaviors
clearly indicate the inapplicability of the simple formula in
describing the magnetoresistance of MgB$_2$
system~\cite{MR_Pallecchi}.

\begin{table}
\caption{Fitting parameters of Fig.~\ref{fig2} using
Eq.~\ref{eq:fit} (All parameters are in SI unit).}
\begin{tabular*}{7cm}{@{\extracolsep{\fill}}ccccccccc}
\hline \hline
$\sigma_{xx}$ &   $a_1$ $(10^{-20})$  &   $a_2$ $(10^{4})$  &   $b_1$ $(10^{24})$  &   $b_2$ $(10^{48})$\\
\hline
$\sigma_1$  &   1.66990    &    10.94632   &   6.76549    &  2.29927 \\
$\sigma_2$  &   1.69337    &    0.29735   &   0.27378    &  0.01295 \\
$\pi_1$     &   4.95528    &    2.20045   &   0.51383    &  0.00998 \\
$\pi_2$     &   1.88788    &    0.32211   &   0.26752    &  0.00269 \\
\hline
$\sigma_{xy}$ &   $c_1$ $(10^{-8})$  &   $c_2$ $(10^{16})$  &   $d_1$ $(10^{24})$  &   $d_2$ $(10^{48})$\\
\hline
$\sigma_1$  &   0.92146    &    2.59031   &   2.90378    &  0.93769 \\
$\sigma_2$  &   0.28948    &    0.41532   &   1.08441    &  0.07650 \\
$\pi_1$     &   0.45035    &    1.46465   &   2.26895    &  0.04918 \\
$\pi_2$     &   0.12735    &    -0.01163  &   0.20542    &  0.00127 \\
\hline \hline
\end{tabular*}
\label{table1}
\end{table}

\begin{figure}
\includegraphics[width=7cm]{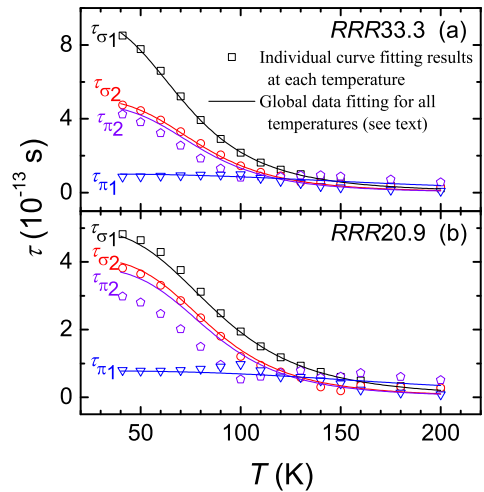}
\caption{(Color online) Temperature dependence of calculated
transport scattering times of the four bands for MgB$_2$ films
RRR33.3 (a), and RRR20.9 (b).} \label{fig3}
\end{figure}

By using the least-square fit to multi-functions~\cite{LS}, we then
fit the experimental data $\sigma_{xx}(B)$ and $\sigma_{xy}(B)$
(derived from the resistivity by
$\sigma_{xx}\equiv\rho_{xx}/(\rho_{xx}^2+\rho_{xy}^2)$ and
$\sigma_{xy}\equiv\rho_{xy}/{(\rho_{xx}^2+\rho_{xy}^2)}$) at each
temperature to Eq.~\ref{eq:fit} by adjusting the four electron
scattering times. The maximum range of $B\tau$ in the fitting are
plotted by the thick solid lines as shown in Fig.~\ref{fig2}.
Figure~\ref{fig3} shows the temperature dependence of four
scattering times for samples RRR33.3 and RRR20.9. Note that the
fitting becomes less reliable at high temperature ($T>100
\mathrm{K}$), as the nonlinearity of the magneto-resistivity becomes
weaker.

To extend our fitting to higher temperature, we adopt a global
fitting approach: Instead of fitting an individual curve at a time,
we simultaneously fit all the curves for all temperatures. To do
that, we assume
$1/\tau^{(n)}=1/\tau_\mathrm{imp}^{(n)}+1/\tau_{e\mathrm{-ph}}^{(n)}\;(T)$,
where $\tau_\mathrm{imp}^{(n)}$ is the scattering time from
impurities and is $T$-independent, and
$\tau_{e\mathrm{-ph}}^{(n)}(T)$ is due to the \textit{e}-ph
coupling, and can be modeled as
$1/\tau_{e\mathrm{-ph}}^{(n)}\;(T)=\alpha_n T^{\beta_n}$.   The
results are shown by the solid lines in Fig.~\ref{fig3}. We find
that both approaches coincide well.

\begin{figure}
\includegraphics[width=8cm]{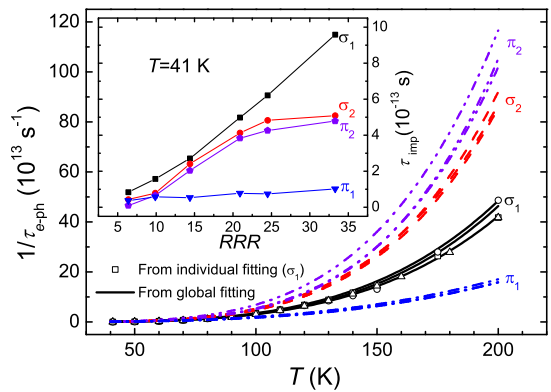}
\caption{(Color online) Temperature dependence of
$1/\tau_{e\mathrm{-ph}}$ of four bands for three different samples
RRR33.3, RRR24.5, and RRR20.9. The inset shows the correlation
between the scattering time from the impurities and $RRR$ for seven
different samples. } \label{fig4}
\end{figure}

We can separate the $T$-independent impurity contribution with the
$T$-dependent \textit{e}-ph contribution. In Fig.~\ref{fig4}, we
plot the temperature dependence of $1/\tau_{e\mathrm{-ph}}$ of three
clean samples RRR33.3, RRR24.9, and RRR20.9.  The  values from all
the three samples are close to each other, indicating the
\textit{e}-ph coupling are not strongly affected by the impurities.
The exponents $\beta_n$ for the temperature dependence of
$1/\tau_{e\mathrm{-ph}}$ are found to be 3.80 ($\sigma_1$), 4.15
($\sigma_2$), 4.22 ($\pi_1$), and 3.12 ($\pi_2$), respectively. This
is in accordance with the expectation of theory~\footnote{For the
simple metals, the low temperature \textit{e}-ph transport
scattering rate follows the so-called ``Bloch $T^5$ law''. However,
for systems with complex Fermi surfaces or small Fermi pockets, the
back-scattering of electron may not require large momentum transfer,
and the transport scattering rate will be proportional to $T^3$.  In
general, the exponent should be in the range of $3$-$5$. See
Ref.~\cite{SolidState}.}. The results clearly show that the $\sigma$
bands have stronger \textit{e}-ph coupling than the $\pi_1$ band, in
agreement with the theoretical calculation~\cite{Monni}.
Surprisingly, the \textit{e}-ph coupling in the $\pi_2$ band is also
very strong. The difference between the \textit{e}-ph coupling in
the two $\pi$ bands can be understood by our \textit{ab initio}
calculation: in the fitting range shown in Fig.~\ref{fig2}, the
magneto-transport of $\pi_2$ band is dominated by the hills of its
Fermi surface, while that of the $\pi_1$ band is dominated by the
belly (see insets of Fig.~\ref{fig2}). Momentum changes to
back-scatter an electron in the hills of $\pi_2$ band, as well as in
the two sigma ones, are much smaller than that in the $\pi_1$ band.
As a result, phonon is much more effective in scattering electrons
in $\pi_2$ and two $\sigma$ bands than in $\pi_1$. The intriguing
result cannot be revealed without our fully band-resolved method.

The inset to Fig.~\ref{fig4} shows the dependence of the impurity
scattering rate on RRR. It shows that disorder affects the impurity
scattering much more strongly in the $\sigma$ bands and $\pi_2$ band
than in the $\pi_1$ band. The calculation results show that for the
dirty sample RRR6.4, the scattering times for the four bands are
comparable.

In conclusion, we have observed large magnetoresistance and
nonlinear Hall effect in clean MgB$_2$ films, from which we have
calculated the scattering times of each one of the four bands, in
the aid of theory based on the Boltzmann equation and \emph{ab
initio} calculations. Surprisingly, the $\pi_2$ band seems to be
similar as the two $\sigma$ bands, i.e., has a large scattering time
in pure samples at low temperature, and also has a strong
\textit{e}-ph coupling. And the electron-phonon scattering is much
weaker in the $\pi_1$ band than other three bands, making the
$\pi_1$ band the least scattered at high temperatures. This may be
caused by the different structure of the Fermi surface of this band.
Disorder modifies the multiband electron scattering much in the same
way as the phonon, reducing the scattering times of the two $\sigma$
bands and $\pi_2$ band to be comparable to the $\pi_1$ band. This
detailed knowledge on the band-resolved scattering rates allow to
have deeper insight on the properties of MgB$_2$ than using the
two-band approximation.

We thank A. Gurevich, I. I. Mazin, M. Putti, C. Ferdeghini for
critical remarks and helpful discussions. This work is supported by
the Natural Science Foundation of China, the MOST Project (Nos.
2006CB601000, 2006CB921802, 2006CB921300, 2007CB925000), the
Knowledge Innovation Project of Chinese Academy of Sciences
(ITSNEM). The work at Penn State is supported by NSF under Grant
Nos. DMR-0306746 (X. X. X.), DMR-0405502 (Q. L.), and DMR-0514592
(Z. K. L. and X. X. X.), and by ONR under Grant No. N00014-00-1-
0294 (X. X. X.). Work in Italy is supported by MIUR, PRIN2006021741,
and PON-CyberSar (COSMOLAB).

$^*$Corresponding authors: experiment and data analysis:
hhwen@aphy.iphy.ac.cn (H. H. Wen); theory: jrshi@aphy.iphy.ac.cn (J.
R. Shi).

\end{document}